\documentclass[pre,twocolumn,showpacs]{revtex4}
\usepackage{amsmath,amsfonts,graphicx}

\newcommand\edge[2]{\langle{#1{}#2}\rangle}

\begin{document}

\title{Alternative determinism principle for topological analysis of
  chaos} \author{Marc Lefranc} \affiliation{Laboratoire de Physique
  des Lasers, Atomes, Mol\'ecules,\\ UMR CNRS 8523, Centre d'\'Etudes et
  de Recherches Lasers et Applications, Universit\'e des Sciences et
  Technologies de Lille, F-59655 Villeneuve d'Ascq, France}

\date{August 21, 2006}

\begin{abstract}
  The topological analysis of chaos based on a
  knot-theoretic characterization of unstable periodic orbits has
  proved a powerful method, however knot theory can only be applied to
  three-dimensional systems.  Still, the core principles upon which
  this approach is built, determinism and continuity, apply in any
  dimension. We propose an alternative framework in which these
  principles are enforced on triangulated surfaces rather than curves
  and show that in dimension three our approach numerically predicts
  the correct topological entropies for periodic orbits of the
  horseshoe map.
\end{abstract}


\maketitle

Chaotic behavior results from the interplay of two geometrical
processes in state space: \emph{stretching} separates neighboring
trajectories while \emph{squeezing} maintains the flow within a
bounded region~\cite{Ott93,topchaos}. A topological analysis has been
developed to classify the ways in which stretching and squeezing can
organize a chaotic attractor~\cite{Mindlin90a,Mindlin91a,topchaos}. It
relies on a theorem stating that unstable periodic orbits (UPO) of a
chaotic three-dimensional (3D) flow can be projected onto a 2D
branched manifold (a \emph{template}) without modifying their knot
invariants~\cite{Birman83a}. In this method, UPO extracted from an
experimental time series are characterized by the simplest template
compatible with their topological
invariants~\cite{Mindlin90a,Mindlin91a,topchaos}.

However this approach can only be applied to 3D attractors: in higher
dimensions, all knots can be deformed into each other. Although other
topological methods are applicable to higher
dimensions~\cite{Froyland01,Day04}, extending template analysis is
still desirable because it provides a different information. A first
step to overcome the 3D limitation is to recognize that knot theory is
not a necessary ingredient but simply a convenient way to study how
two fundamental properties, determinism and continuity, constrain
trajectories in phase space. It is because two trajectories cannot
intersect that the knot type of a 3D periodic orbit is well defined
and is not modified as the orbit is deformed under control parameter
variation.

In this paper, we note that a dimension-independant formulation of
determinism is orientation preservation and propose an approach where
it is enforced on a representation of the dynamics in a triangulation
of periodic points. In dimension three, an explicit formalism is
easily constructed, and we find that it numerically predicts the
correct entropies for periodic orbits of the horsehoe map. The entropy
of a periodic orbit is an invariant defined as the minimal topological
entropy~\cite{Walters} of a flow containing this
orbit~\cite{Boyland94:_topol,bestvina95:_train}; a positive-entropy
orbit is a powerful indicator of
chaos~\cite{Mindlin91a,amon04:_topol,Thiffeault05}. This result
suggests that a key ingredient for constructing a knotless template
analysis has been captured, although a proof of validity and an
explicit higher-dimensional extension are still lacking.

%%% end of introduction

\begin{figure}[htbp]
  \centering
  \includegraphics[width=3.375in]{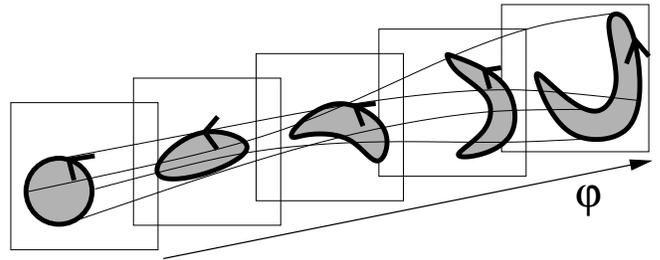}
  \caption{Under the action of the flow, volume elements of
    Poincar\'e sections and their boundaries are stretched and
    squeezed but retain their orientation, as illustrated here for 2D
    section planes.   } 
  \label{fig:orientation}
\end{figure}

We now detail our approach. A first step is to replace the requirement
of non-intersecting curves by a geometrical problem that adapts
naturally to phase spaces of any dimension. It has been suggested to
exploit the rigid structure of invariant manifolds of
UPO~\cite{Mindlin91a,Mindlin97:_tori_klein_bottl}. Here, we note that
when a volume element $V$ of a $d$-dimensional phase space is advected
by a deterministic flow $\Phi_t$, the image $\Phi_t(\partial
V)$ of its boundary cannot display
self-intersections: at any time $t$, its interior and its exterior
remain distinct, as with a droplet in a fluid flow. A technical
formulation of this property is that volume orientation is preserved
by the dynamics. For simplicity, we consider attractors embedded in
$\mathbb{R}^n \times S^1$ (e.g., forced systems), that can be sliced
into $n$-dimensional Poincar\'e sections parameterized by $\varphi \in
S^1$. Determinism then imposes that boundaries of $n$-dimensional
volume elements of Poincar\'e sections retain their orientation
(Fig.~\ref{fig:orientation}).

\begin{figure}[htbp]
  \centering
  \includegraphics[width=3.375in]{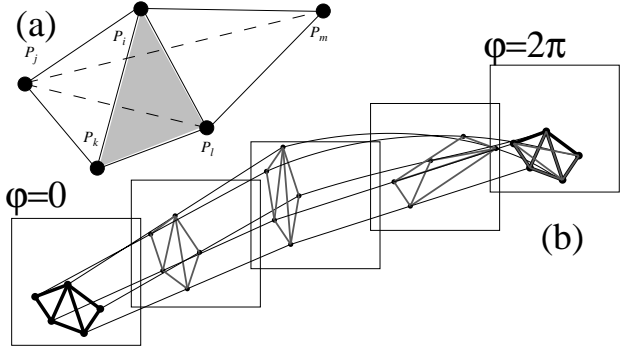}
  \caption{(a) Triangulated space based on periodic points $P_i$ in a
    3D Poincar\'e section. The 2-cell $\langle ikl\rangle$ is shaded.
    (b) The flow induces a mapping of this triangulated space into
    itself, as suggested here for a period-5 orbit embedded in
    $\mathbb{R}^2\times S^1$. }
  \label{fig:simplicial}
\end{figure}

Template analysis must be applicable to UPO extracted from
experimental signals, and thus can only rely on the phase space
trajectory of a period-$p$ orbit. 
Thus, we represent the dynamics in a triangulated space whose
nodes are $p$ periodic points $P_i$ in a Poincar\'e section, with
$P_{i+1}=F(P_i)$, $F$ being the return map. In this space, points
$P_i$ are 0-cells, line segments $\langle P_i,P_j\rangle\equiv \langle
ij\rangle$ joining two points are 1-cells, triangles $\langle
P_i,P_j,P_k\rangle \equiv \langle ijk\rangle$ are 2-cells, etc
(Fig.~\ref{fig:simplicial}a).  Similar concepts have been used
in~\cite{Sciamarella99:_topol} to analyze the static
structure of 
an attractor, but we focus here on the dynamics. We denote by $S_m$ the
set of collections of contiguous $m$-cells, which are the analogs of
$m$-dimensional surfaces in 
the original phase space. As Poincar\'e sections are swept, periodic
points move in the section plane and so do the $m$-cells
attached to them (Fig.~\ref{fig:simplicial}b). The dynamics induced in
$S_m$ should reflect that of $m$-dimensional phase-space surfaces
under action of the chaotic flow, and in particular be organized by
the same stretching and squeezing mechanisms.

A dynamics in the triangulated space is specified by maps $F_m:S_m \to
S_m$ acting on collections of contiguous $m$-cells. Since the original
return map $F$ sends nodes to nodes but not facets to
facets, the $F_m$ are not restrictions of $F$ for $m>0$. However we
require them to mimic $F$ in the following way: they should be
invertible, satisfy determinism and result from a continuous
deformation of facets, just as $F$ is
a continuous deformation of identity.
The $F_m$ should also satisfy $\partial F_m(\Sigma) =
F_{m-1}(\partial \Sigma)$ where $\partial$ is the boundary operator.
As we see below, facets
are not necessarily trivially advected between sections because
degeneracies occur, at which action must be taken to preserve
orientation.

\begin{figure}[htbp]
  \centering
  \includegraphics[width=3.375in]{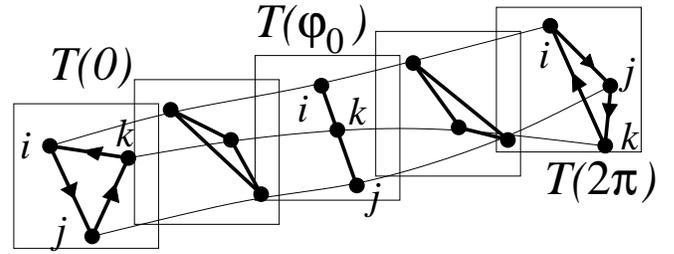}
  \caption{As three points move in the section plane when $\varphi$ is
  increased, the triangle 
    they form can change its orientation.}
  \label{fig:inversion}
\end{figure}

We now specialize to the 3D case. The volume element of a
triangulated set of periodic points in a 2D Poincar\'e section is a
triangle (2-cell) based on three periodic points $P_i$, $P_j$, $P_k$.
Let $P_i(\varphi)$ be the position of $P_i$ in
section $\varphi$, with $P_i(0) = P_i$ and $P_i(2\pi)
= P_{i+1}$. The natural evolution of $T=\langle
P_i,P_j,P_k \rangle$ as $\varphi$ increases is
\begin{equation}
  T(\varphi)=\langle P_i(\varphi),P_j(\varphi),P_k(\varphi) \rangle,
  \label{eq:naive}
\end{equation}
which would lead to a trivial induced return map $F_2(T) = T(2\pi) =
\langle P_{i+1},P_{j+1},P_{k+1}\rangle$ if~\eqref{eq:naive} was
uniformly valid as a 2-cell.  However, it is common that at some
$\varphi=\varphi_0$, one of the three points [say $P_k(\varphi)$]
passes between the two others, thereby changing the orientation of the
candidate 2-cell $T(\varphi)$ given by~\eqref{eq:naive}
(Fig.~\ref{fig:inversion}). As emphasized above, this is strictly
forbidden by determinism, and we must thus modify the representation
of the dynamics. It turns out that this problem has a simple solution.

The degenerate triangle $T(\varphi_0)$ in Fig.~\ref{fig:inversion} is
like a flattened balloon whose boundary splits into two
superimposed sides with opposing outer normals. Determinism is
violated when these two sides cross each other so that interior and
exterior, defined with respect to outer normal, seem to be exchanged.
However, the experimental data only constrain node motion, from which
the facet dynamics is interpolated. To preserve determinism, we force
the two opposing sides not to cross by swapping them at degeneracy,
thereby canceling the inversion.

\begin{figure}[htbp]
  \centering
  \includegraphics[width=3.375in]{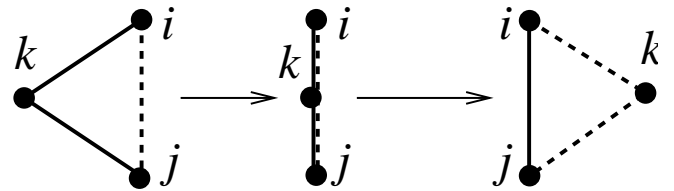}
  \caption{A triangle is inverted as $P_k$ passes between $P_i$
    and $P_j$. Identifying the solid (resp., dashed) paths in the
    initial and end configurations leads to
    substitution~\eqref{eq:subs}.}
  \label{fig:subs}
\end{figure}

This prescription is illustrated in Fig.~\ref{fig:subs} where the two
opposing sides at triangle degeneracy are represented as a solid and a
dashed line. The key point is that we construct the edge dynamics so
that the left (solid line) and right (dashed line) sides remain at the
left and right, respectively. Since the left (resp., right) side
consists of itinerary $\edge{i}{k}+\edge{k}{j}$ (resp., $\edge{i}{j}$)
before degeneracy and of itinerary $\edge{i}{j}$ (resp.,
$\edge{i}{k}+\edge{k}{j}$) after degeneracy, their relative position
is preserved by applying the following dynamical rule in $S_1 =
S_{n-1}$  at triangle inversion:
\begin{subequations}
  \label{eq:subs}
  \begin{eqnarray}
  \edge{i}{j} &\to& \edge{i}{k}+\edge{k}{j}\label{eq:subs_stretch}\\ 
   \edge{i}{k}+\edge{k}{j} &\to&    \edge{i}{j} \label{eq:subs_squeeze}
\end{eqnarray}
\end{subequations}
These rules also apply to reverse paths (e.g.,
$\edge{j}{i}\to\edge{j}{k}+\edge{k}{i}$).  Note that
$\partial T = \partial \langle ijk \rangle =
\edge{i}{j}+\left(\edge{j}{k}+\edge{k}{i}\right)$ is mapped
by~\eqref{eq:subs} to
$\left(\edge{i}{k}+\edge{k}{j}\right)+\edge{j}{i}= \partial \langle
ikj \rangle$. The permutation compensates for triangle inversion so
that orientation of $\partial T$, and hence determinism, is preserved.

Itineraries visiting edges
$e_{ij}=\edge{i}{j}$ in a given order are represented by words in a
language $\mathcal{A}^*$ over alphabet $\mathcal{A}=\{e_{lm}\}$, and
~\eqref{eq:subs} by an operator $\sigma_{ij}^k$ that in
each word $w$ replaces the letter $e_{ij}$ by the string
$e_{ik}e_{kj}$ and $e_{ik}e_{kj}$ by $e_{ij}$ [hence
$(\sigma_{ij}^k)^2=1$]. For example,
\begin{displaymath}
  \sigma_{ij}^k\,
  e_{kl}e_{li}\overline{e_{ij}}e_{jl}e_{li}\underline{e_{ik}e_{kj}}
  \overline{e_{ji}}\ldots
  =
  e_{kl}e_{li}\overline{e_{ik}e_{kj}}e_{jl}e_{li}\underline{e_{ij}}
  \overline{e_{jk}e_{ki}}\ldots
\end{displaymath}
The $\sigma_{ij}^k$ generate a non-trivial dynamics, as the image of
an itinerary depends on how periodic points rotate around each other.
This simple dynamics faithfully reflects that of the flow around the
periodic orbit, as we show by computing the entropy of
the orbit.

\begin{figure}[htbp]
  \centering
  \includegraphics[width=3.375in]{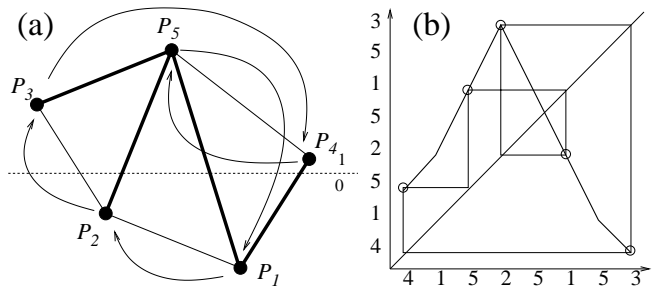}
  \caption{(a) Periodic points of the horseshoe orbit 00111 and their
    schematic trajectory in section plane. Bold lines indicate edges
    involved in~\eqref{eq:00111}. (b) Path
    $P_4 P_1 P_5 P_2 P_5 P_1 P_5 P_3$ folds onto itself under action
    of induced return map $F_1$. The unimodal map obtained has 00111
    as a periodic orbit.}
  \label{fig:00111}
\end{figure}

From the motion of periodic points $P_i(\varphi)$ in the section plane
as $\varphi$ is swept, a list of $l$ triangle inversions $\sigma_{i_m
  j_m}^{k_m}$ is obtained, from which we build an induced return map
that transforms a word $w\in\mathcal{A}^*$ into another word $w'$ as:
\begin{equation}
  \label{eq:operation}
  F_1 : w \to w' =  N \sigma_{i_l j_l}^{k_l}\cdots\sigma_{i_2 j_2}^{k_2}
  \sigma_{i_1 j_1}^{k_1} \,w,
\end{equation}
where $N\, e_{ij}\cdots =e_{(i+1)(j+1)}\cdots$. Consider periodic
orbit $00111$ of a suspension of the standard horseshoe map equipped
with the usual symbolic coding~\cite{topchaos}
(Figs.~\ref{fig:simplicial}b and \ref{fig:00111}a). We find that as
points gradually move in the section plane from their initial location
to that of their image under the return map, triangle inversions occur
when point $4$ successively crosses the four edges $e_{15}$, $e_{13}$,
$e_{25}$ and $e_{23}$. Thus the induced return map for edge
itineraries is $F_1=N
\sigma_{23}^4\sigma_{25}^4\sigma_{13}^4\sigma_{15}^4$. For example,
\begin{displaymath}
e_{15}\xrightarrow{\sigma_{15}^4}
e_{14}e_{45} \xrightarrow{\cdots} e_{14}e_{45} \xrightarrow{N}
e_{25}e_{51} = F_1(e_{15}),
\end{displaymath}
while edges not crossed by point 4 are trivially modified (e.g.,
$e_{14}\xrightarrow{N} e_{25}$). This leads to the closed rule set
\begin{equation}
  \label{eq:00111}
  e_{14} \to e_{25},\;
  e_{15} \to e_{25}\,e_{51},\;
  e_{25} \to e_{35}\,e_{51},\;
  e_{35} \to e_{41}
\end{equation}
for edges in the invariant set of $F_1$.
Table~\ref{tab:iter} displays iterates $F_1^m(e_{15})$
computed using~\eqref{eq:00111}. Their length $|F_1^m(w)|$ diverges
exponentially as $m\to\infty$, indicating that trajectories in the
neighborhood of the orbit are continuously stretched apart by the
flow.  The growth rate:
\begin{equation}
  \label{eq:entropy}
  h(P) = \lim_{m\to\infty} \frac{\ln |F_1^m(w)|}{m}
\end{equation}
is obtained as the logarithm of the leading eigenvalue of the
transition matrix $(M_{ee'})$, whose entries count occurences of edge
$e'$ or of its reverse in $F_1(e)$ given by~\eqref{eq:00111}. Here,
$h(00111)\sim0.5435$. Table~\ref{tab:iter} also shows that
$F_1^{kp}(w)$ ($p$ is the orbit period) converges to an infinite word
$w_\infty$ satisfying $F_1^p(w_\infty)=w_\infty$, which is the analog
of the infinitely folded unstable manifold of the periodic orbit.

\begin{table}[htbp]
  \centering
  \begin{tabular}[c]{ll}
    \hline
    $m$& Itinerary of $F_1^m(e_{15})$\\\hline
    
    $0$ & $(15)$ \\
    $1$ & $(251)$\\
    $2$ & $(35152)$\\
    $3$ & $(41525153)$\\
    $4$ & $(5251535152514)$\\
    $5$ & $(1535152514152515351525)$\\
    $6$ &  $(2514152515351525251535152514152515351)$\\
    $10$ & $(1535152514152515351525251535152525153515\ldots)$\\
    $15$ & $(1535152514152515351525251535152525153515\ldots)$\\
    $100$ & $(1535152514152515351525251535152525153515\ldots)$\\
    \hline
  \end{tabular}
  \caption{A few iterates
    $F_1^m(e_{15})$ are given by their itinerary between
    periodic points [e.g., (35152) denotes the path
    $e_{35}e_{51}e_{15}e_{52}$].} 
  \label{tab:iter}
\end{table}

The growth rate $h(P)$ is expected to be the entropy $h_T(P)$ of orbit
$P$, defined as the minimal topological entropy~\cite{Walters} of a
map containing $P$~\cite{Boyland94:_topol}. Indeed, a piecewise linear
map containing $P$ with $(M_{ee'})$ as Markov transition matrix can be
constructed and has entropy $h(P)$, thus $h_T(P)\le h(P)$. Conversely,
$h(P)\le h_T(P)$, as $h(P)$ is the minimal growth rate of the geometric
length of curves passing through periodic points $P_i$ and cannot
be larger than the topological entropy of a map containing $P$, which
is the supremum of stretching rates over curves in the
plane~\cite{Newhouse93}.

For a typical orbit, unlike in~\eqref{eq:00111}, there are paths in
the $F_1$-invariant set that trigger a ``squeezing''
rule~\eqref{eq:subs_squeeze}, as for example $e_{16}e_{67} \to e_{17}$
for horseshoe orbit $0010111$. Then $F_1(e_{16}e_{67})\neq
F_1(e_{16})F_1(e_{67})$ and the transition matrix cannot be used for
entropy computations, although estimates can still be obtained by
direct iteration. In all examples we considered, enlarging the
alphabet by recoding contracting paths as basis edges (e.g.,
$e_{167}\equiv e_{16}e_{67}$) and applying other recodings
required for consistency allowed us to rewrite $F_1$ as an ordinary
substitution like~\eqref{eq:00111}. For example, the induced return map
for horseshoe orbit 0010111 can be rewritten as ($e_{ijk}\equiv
e_{ij}e_{jk}$):
\begin{displaymath}
  \label{eq:0010111}
  \begin{array}[c]{l}
  e_{14} \to e_{25},\;
  e_{15} \to e_{257}\,e_{76},\;
  e_{17} \to e_{257}\,e_{71},\;
  e_{25} \to e_{37}\,e_{76},\\
  e_{37} \to e_{41},\;
  e_{67} \to e_{71},
  e_{167} \to e_{25}\,e_{51},\;
  e_{257} \to  e_{37}\;e_{761}
\end{array}
\end{displaymath}
Besides $e_{167}$, basis path
$e_{257}$ was introduced because its image overlaps $e_{167}$.
A transition matrix can then be obtained, with entropy $h(0010111)\sim
0.4768$.

For all 746 periodic orbits of the horseshoe map up to period 12, we
have compared growth rate~\eqref{eq:entropy} with topological entropy
obtained by the train-track
algorithm~\cite{bestvina95:_train,Boyland94:_topol,Hall:_train}. As
illustrated in Table~\ref{tab:entropies}, we found \emph{agreement to
  machine precision in each instance}. This strongly suggests that in
3D, our approach is equivalent to the train track approach.
Qualitative properties of chaos are also reproduced: the dynamics is
deterministic (by construction), invertible, and the stretching and
squeezing processes are described in a symmetrical way.

\begin{table}
\newcommand\zo{\;_1^0}
\begin{tabular}{lcc|lcc}
  \hline
   Orbit& This work & TTA& Orbit & This work & TTA\\\hline
   $01101\zo1$&   $0.4421$&   $0.4421$ &
   $00010\zo1$&   $0.3822$&   $0.3822$ \\
   $001011\zo1$&   $0.3460$&   $0.3460$ &
   $000101\zo1$&   $0.5686$&   $0.5686$ \\
   $00101\zo1$&   $0.4768$&   $0.4768$ &
   $0001\zo1$&   $0.6329$&   $0.6329$ \\
   $001010\zo1$&   $0.4980$&   $0.4980$ &
   $000111\zo1$&   $0.5686$&   $0.5686$ \\
   $001\zo1$&   $0.5435$&   $0.5435$ &
   $00011\zo1$&   $0.3822$&   $0.3822$ \\
   $001110\zo1$&   $0.4980$&   $0.4980$ &
   $000010\zo1$&   $0.4589$&   $0.4589$ \\
   $00111\zo1$&   $0.4768$&   $0.4768$ &
   $00001\zo1$&   $0.6662$&   $0.6662$ \\
   $001111\zo1$&   $0.3460$&   $0.3460$ &
   $000011\zo1$&   $0.4589$&   $0.4589$ \\
   $001101\zo1$&   $0.4980$&   $0.4980$ &
   $000001\zo1$&   $0.6804$&   $0.6804$\\          \hline
  \end{tabular}
  \caption{Topological entropies of
  positive-entropy horseshoe orbits up to period 8 obtained with the approach
  described here and  with the
  train-track algorithm (TTA).} 
\label{tab:entropies}
\end{table}

Remarkably, we note that while transformations~\eqref{eq:operation}
are \emph{invertible}, the asymptotic dynamics is \emph{singular}.
Consider the itinerary $w_0=F^3(e_{15})=(41525153)$ in
Table~\ref{tab:iter}, which is the shortest subpath of $w_\infty$
visiting the four edges in~\eqref{eq:00111}. As
Fig.~\ref{fig:00111}(b) shows, the image
$F_1(w_0)=(5251535152514)=(525153)+(35152514)$ consists of a subpath
of $w_0$ concatenated with a reverse copy of $w_0$: \emph{this path is
  folded onto itself by a singular one-dimensional map}. The same
property holds for all subsequent iterates $F^m(e_{15})$, hence for
the infinite word $w_\infty$. This reflects that associated to an
invertible return map (e.g., H\'enon map), there exists an underlying
lower-dimensional noninvertible map (e.g., logistic map) describing
the dynamics along the unstable manifold, a keystone of the
Birman-Williams construction~\cite{Birman83a,topchaos}. Note that the
symbolic name 00111 can be recovered directly from
Fig.~\ref{fig:00111}(b) using the usual coding for orbits of 1D maps.
This makes the new formalism promising for using topological
information to construct global symbolic codings as
in~\cite{Plumecoq2000a}. How segments along $w_0$ are folded over each
other and how neighboring cells are squeezed provide us with a
combinatorial description of stretching and folding that could be used
to determine the simplest template carrying the periodic orbit
studied.

To conclude, we have proposed that orientation preservation is a more
general formulation of determinism than non-intersection of
trajectories. In three dimensions we find that enforcing it on a
triangulation of periodic points induces a nontrivial dynamics on
paths along periodic periodic points. More precisely, a path map $F_1$
is constructed by: (i) following triangles advected by the flow as one
rotates around the attractor, (ii) restoring orientation at each
triangle inversion by exchanging opposing sides via
transformations~\eqref{eq:subs}. When paths in the $F_1$-invariant set
do not experience contraction, entropy is obtained from an transition
matrix indicating how elementary edges in the invariant set are mapped
among themselves. Otherwise, new basis paths must be introduced to
account for contraction.  A promising result is that despite its
simplicity this formalism numerically predicts the correct entropies
for periodic orbits of the horseshoe map. Preliminary calculations
also suggest that it leads to a combinatorial description of the
folding of the invariant unstable manifold over itself, yielding
information about the symbolic dynamics of the orbit. It now remains
to prove the validity of the approcach in 3D and to try to extend it
to higher dimensions.

This work grew out of innumerable discussions with R. Gilmore. I thank
T. Hall, J. Los and F. Gautero for helpful explanations about train
tracks, and M. Nizette, T. Tsankov, J.-C. Garreau, C. Szwaj and S.
Bielawski for a careful reading of this manuscript. CERLA is supported
by the Minist\`ere charg\'e de la Recherche, R\'egion Nord-Pas de
Calais and FEDER.


\begin{thebibliography}{10}

\bibitem{Ott93}
E. Ott, {\em Chaos in Dynamical Systems} (Cambridge University Press,
  Cambridge, 1993).
  
\bibitem{topchaos} R. Gilmore, Rev. Mod. Phys. {\bf 70}, 1455 (1998).
  R. Gilmore and M. Lefranc, {\em The Topology of Chaos} (Wiley, New
  York, 2002).

\bibitem{Mindlin90a}
G.~B. Mindlin {\it et~al.}, Phys. Rev. Lett. {\bf 64},  2350  (1990).

\bibitem{Mindlin91a}
G.~B. Mindlin {\it et~al.}, J. Nonlinear Sci. {\bf 1},  147  (1991).

\bibitem{Birman83a}
J.~S. Birman and R.~F. Williams, Topology {\bf 22},  47  (1983).

\bibitem{Froyland01}
G. Froyland, O. Junge, and G. Ochs, Physica D {\bf 154},  68  (2001).

\bibitem{Day04} S. Day, O.Junge and K. Mischaikow, SIAM J. Appl. Dyn.
  Sys. {\bf 2}, 117 (2004).


\bibitem{Walters} P. Walters, {\em An Introduction to Ergodic Theory}
  (Springer, New York, 2000)
  
\bibitem{Boyland94:_topol}
P. Boyland, Topology Appl. {\bf 58},  223  (1994).

\bibitem{bestvina95:_train}
M. Bestvina and M. Handel, Topology {\bf 34},  109  (1995).

\bibitem{amon04:_topol}
A. Amon and M. Lefranc, Phys. Rev. Lett. {\bf 92},  094101  (2004).

\bibitem{Thiffeault05}
J.-L. Thiffeault, Phys. Rev. Lett. {\bf 94},  084502  (2005).




\bibitem{Mindlin97:_tori_klein_bottl}
G.~B. Mindlin and H.~G. Solari, Physica D {\bf 102},  177  (1997).

\bibitem{Sciamarella99:_topol}
D. Sciamarella and G.~B. Mindlin, Phys. Rev. Lett. {\bf 82},  1450  (1999).

\bibitem{Newhouse93}
S.~E. Newhouse and T. Pignataro, J. Stat. Phys. {\bf 72},  1331  (1993).

\bibitem{Hall:_train}
T. Hall, \textsc{trains}, software available from
  {http://www.liv.ac.uk/\\maths/PURE/MIN\_SET/CONTENT/members/T\_Hall.html}.

\bibitem{Plumecoq2000a}
J. Plumecoq and M. Lefranc, Physica D {\bf 144},  231  (2000).


\end{thebibliography}
\end{document}